\documentclass{elsart}
\usepackage{graphicx}

\def\Journal#1#2#3#4{{#1} {\bf #2}, (#4) #3}

\def\NIMA{{Nucl. Instr. Meth.} A}
\def\NIMB{{Nucl. Instr. Meth.} B}
\def\NPA{{Nucl. Phys.} A}

\def\PLB{{Phys. Lett.}  B}
\def\PRL{Phys. Rev. Lett.}

\def\ADV{Adv. Space Res.}
\def\APP{Astropart. Phys.}

\newcommand{\mc}{{\footnotesize MC}}
\newcommand{\beam}{{\footnotesize BEAM}}
\newcommand{\cfg}[1]{{\footnotesize CFG{#1}}}

\begin{document}
\begin{frontmatter}
  \title{
	Measurement of low-energy antiproton detection efficiency 
	in BESS	below 1 GeV 
    }

  \author[Tokyo]{Y.~Asaoka\thanksref{author}},
  \author[KEK]{K.~Yoshimura},
  \author[KEK]{T. Yoshida},
  \author[Tokyo]{K.~Abe},
  \author[Tokyo]{K.~Anraku},
  \author[Tokyo]{M.~Fujikawa},
  \author[Tokyo]{H.~Fuke},
  \author[Tokyo]{S.~Haino},
  \author[Tokyo]{K.~Izumi},
  \author[Kobe]{T.~Maeno},
  \author[KEK]{Y.~Makida},
  \author[Tokyo]{N.~Matsui},
  \author[Tokyo]{H.~Matsumoto},
  \author[Tokyo]{H.~Matsunaga\thanksref{tsukuba}},
  \author[Tokyo]{M.~Motoki\thanksref{tohoku}},
  \author[Kobe]{M.~Nozaki},
  \author[Tokyo]{S.~Orito\thanksref{orito}},
  \author[Tokyo]{T.~Sanuki},
  \author[Kobe]{M.~Sasaki},
  \author[Tokyo]{Y.~Shikaze},
  \author[Tokyo]{T.~Sonoda},
  \author[KEK]{J.~Suzuki},
  \author[KEK]{K.~Tanaka},
  \author[Kobe]{Y.~Toki} and
  \author[KEK]{A.~Yamamoto}

  \thanks[author]{Corresponding author. E-mail:
    asaoka@icepp.s.u-tokyo.ac.jp. Tel: +81 3 3815 8384 \& Fax: +81 3
    3814 8806}
  \thanks[tsukuba]{present address: University of Tsukuba, Tsukuba,
    Ibaraki, 305-8571, Japan}
  \thanks[tohoku]{present address: Tohoku University, Sendai, Miyagi, 
    980-8577, Japan}
  \thanks[orito]{deceased.}

  \address[Tokyo]{The University of Tokyo,
    Hongo, Bunkyo, Tokyo, 113--0033 Japan}
  \address[KEK]{High Energy Accelerator Research Organization (KEK),
    Oho, Tsukuba, Ibaraki, 305--0801 Japan}
  \address[Kobe]{Kobe University,
    Rokkodai-cho, Nada, Kobe, 657--8501 Japan}

  \begin{abstract}

An accelerator experiment was performed using a low-energy
antiproton beam to measure antiproton detection efficiency of BESS, a
balloon-borne spectrometer with a superconducting solenoid.
Measured efficiencies showed good agreement with calculated ones
derived from the BESS Monte Carlo simulation based on {\sc geant/gheisha}. 
With detailed verification of the BESS simulation,
the relative systematic error of detection
efficiency derived from the BESS simulation has been determined to be
$\pm$5~\%, compared with the previous estimation of $\pm$15~\% which
was the dominant uncertainty for measurements of cosmic-ray antiproton
flux.

  \end{abstract}
\end{frontmatter}

\section{Introduction}

\begin{figure}[hbt]
  \begin{center}
    \includegraphics[width=8.0cm]{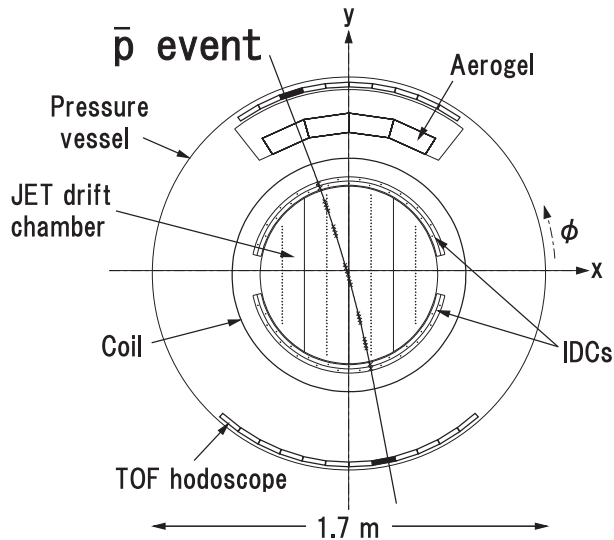}
  \end{center}
  \caption  {Cross-sectional view of the BESS detector showing a
    $\bar{p}$ event.} 
  \label{fig:bess99}
\end{figure}

The BESS spectrometer, shown in Fig.~\ref{fig:bess99}, was
designed~\cite{OR87,YA94} and
developed~\cite{kn:det,YA88,kn:ac,kn:tof} as a high-resolution 
balloon-borne
spectrometer with the capability to search for rare cosmic-rays and
provide various precision measurements of cosmic-ray primaries. 
In spite of many advantages in cylindrical configuration, the adoption 
of solenoidal magnets could not be realized
in previous balloon-borne
spectrometers because of much unavoidable material in particle
passage.
However, a thin superconducting solenoid developed at KEK
\cite{YA94,YA88} enabled us to uniquely adopt this concentric
configuration. 
A uniform magnetic field of 1 Tesla is produced by a thin
superconducting coil~\cite{YA88}, through which particles can pass
without too many interactions.
The magnetic-field region is filled with a tracking detectors
(JET/IDCs), resulting in an acceptance of 0.3 m$^2$sr. 
Tracking is performed by fitting up to 28 hit points in the drift
chambers.
The upper and lower scintillator hodoscopes (TOF)~\cite{kn:tof}
provide time-of-flight and two d$E$/d$x$ measurements.
The instrument also incorporates a threshold-type Cherenkov
counter~\cite{kn:ac} with a silica-aerogel radiator to distinguish
high-energy $\bar{p}$'s from electron and muon backgrounds.

Since 1993 through 2000,
seven balloon flights have been successfully carried out
and more than $2\times10^3$ antiprotons ($\bar{p}$'s) have been
unambiguously detected
\cite{kn:pbar93,kn:pbar95,kn:pbar97,kn:pbar98,kn:pbar00}. 
This has allowed measuring the energy spectrum of cosmic 
$\bar{p}$'s and investigating their origin.
To investigate the origin of low energy $\bar{p}$'s more sensitively,
it is inevitably important to reduce systematic errors as well as
statistical errors of the resultant spectrum.
The dominant source of systematic error in the low energy region ($<$
1~GeV) is the uncertainty in $\bar{p}$ interaction losses in the
instrument. 
In the previous analyses,
we surmised $\pm$15~\% relative error to the
detection efficiency ($\varepsilon$), which is defined as 
\[ \varepsilon = N_{\rm obs}/N_{\rm inc},
\]
where $N_{\rm inc}$ is the number of incidence 
within the acceptance of the detector, and $N_{\rm obs}$ is the number 
of identified particles.
Since $N_{\rm inc}$ cannot be derived directly from flight data, 
the detection efficiency is
evaluated using  
the Monte Carlo simulation (BESS MC) \cite{kn:d-matsu} based on the
{\sc geant/gheisha} code \cite{kn:geant,kn:gheisha}.
The BESS MC incorporates detailed material and detector descriptions
such that realistic detector performance is obtained.
The original {\sc gheisha} code was modified so that experimental data
of $\bar{p}$-nuclei cross sections are reproduced
\cite{kn:pbcprl,p-nuclei}.
However, it is difficult to estimate the systematic error due to
interaction losses because of uncertainties in secondary multiplicity,
angular 
distribution, and detector response;
accordingly, detection efficiency must be directly measured and must
precisely verified to reduce systematic error in the low-energy region
$\bar{p}$ flux.

Considering this, we performed an accelerator beam experiment at the
KEK-PS K2 beam line using a low-energy $\bar{p}$ and proton ($p$)
beam. 
The objectives of the beam experiment were as follows: 
\begin{enumerate}
\item directly measure detection efficiencies for $\bar{p}$'s and $p$'s;
\item examine the BESS MC simulation;
\item reduce systematic error in detection efficiency especially for 
$\bar{p}$'s.
\end{enumerate}
Although the BESS detector has been successively upgraded since the
first successful flight in 1993, 
the basic detector concept is the same, {\it i.e.},
(i) large-acceptance cylindrical configuration with solenoidal magnet,
and (ii) mass identification using the rigidity and velocity
measurements.
Accordingly, the presented beam results can be applied to both past
and future detectors. 

Section~\ref{sec:exp} summarizes experimental setup,
after which Section~\ref{sec:anal} describes incident beam
identification which determines $N_{\rm inc}$.
Measurement of the detection efficiency is then presented in
Section~\ref{sec:rslt}, with a study of simulation results being
discussed in Section~\ref{sec:dsc}. 
Finally, Section~\ref{sec:conc} summarizes the main results and
provides conclusions.    

\section{Experimental Setup}
\label{sec:exp}

\begin{figure}[bhtp]
  \begin{center}
    \includegraphics[width=13.5cm]{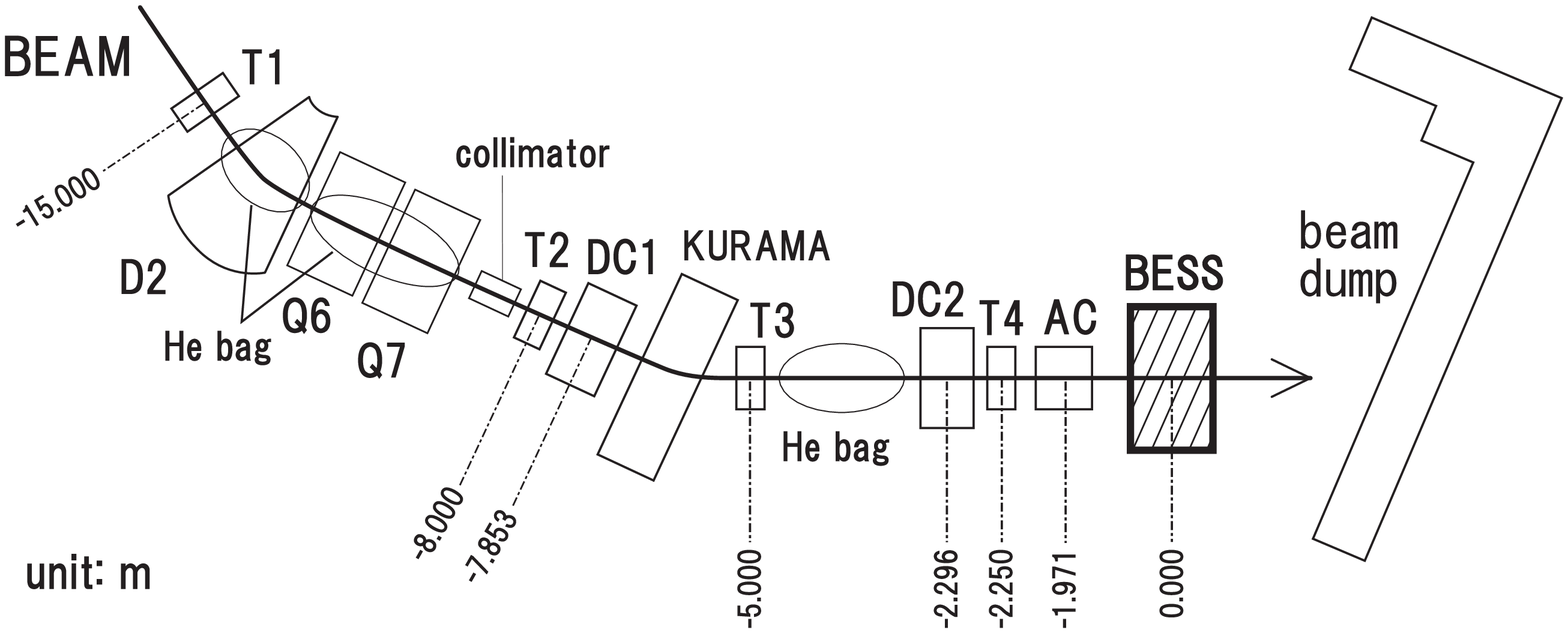}
  \end{center}
  \caption  {
  Experimental setup. 
  T1 -- T4 are 10 mm thick plastic scintillators, having time
  resolutions of 30 -- 40 ps.
  DCs use a mixture of 50 -- 50~\% Ar-ethane gas with a full drift
  length of 1 cm. Each DC has three planes of horizontal wires and
  three planes of vertical wires.
  The spatial resolution is 150 $\mu$m/wire.
  AC uses 8-cm thick aerogel radiator with a refractive index of
  1.03. 
  }
  \label{fig:setup}
\end{figure}

The BESS beam experiment was performed in February 1999
at the KEK-PS K2 beam line which is equipped with an electro-static 
separator~\cite{kn:YA82} to enrich low energy $\bar{p}$'s. 
Figure~\ref{fig:setup} shows a schematic view of the experimental
setup at the down stream of the K2 beam line.
D2 is a dipole magnet, Q6 and Q7 are quadrapole magnets. They
are used for beam transport and focusing.
KURAMA is another dipole magnet to analyze momentum of the incident
particles. 
The BESS detector was rotated circumferentially 70$^\circ$ in
$r$-$\phi$ plane (a plane perpendicular to the axis of the solenoid)
such that it was suitably 
positioned in the beam line for proper (from top to bottom) beam
incidence. 
To identify incident particles and to reject interacted events, we
placed four trigger counters (T1--T4), two drift chambers (DC1 and
DC2), and an aerogel Cherenkov counter (AC) in the beam line (for
details, see figure caption).

Data from the BESS detector and beam line detectors are collected
using the BESS data acquisition system. 
To obtain $N_{\rm inc}$, 
instead of using the BESS standard trigger (coincidence between upper
and lower TOF counters), the trigger was generated by T1 \& T2 \& T3
\& T4 \& ${\overline {\rm AC}}$.
Note that ${\overline {\rm AC}}$ remarkably improves the
$\bar{p}$/$\pi$ ratio.

\begin{figure}[hbt]
  \begin{center}
    \includegraphics[width=4.0cm]{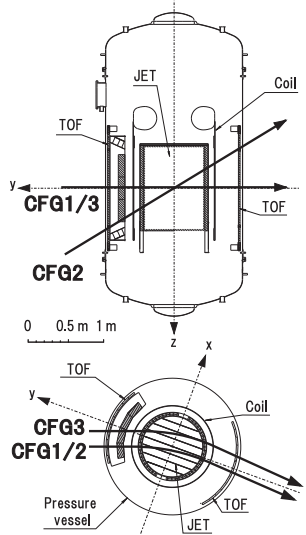}
  \end{center}
  \caption {Cross-sectional views of the BESS detector with incident 
    $\bar{p}$ beams for three configurations (CFG1--3). 
    Using CFG1 as a reference, CFG2 is the case in which the incident
    beam has an angle of cos$\theta = 0.915$, where $\theta$ is
    defined as the angle between the beam and $y$-axis in the $y$-$z$
    plane. CFG3 is the case in which the incident beam passes through
    a different region of the central tracking system in $r$-$\phi$
    plane. The distance of incident position between CFG1 and CFG3 is
    153 mm along the $x$-axis. 
}
  \label{fig:posi}
\end{figure}

Data were collected for three different detector configurations
to represent the typical incidence of cosmic-ray particles in terms of
the amount of material and penetrated region.
Three configurations, \cfg{1--3}, are shown in
Fig.~\ref{fig:posi} together with typical $\bar{p}$ trajectories.
The kinetic energy of incident particles at the BESS top of
instrument ($E_{\rm TOI}$) ranges from 0.1 to 1~GeV (0.4 to 1~GeV) for 
$\bar{p}$'s ($p$'s). 
The BESS detector cannot be rotated more than 70 $^\circ$
due to a constraint of the internal structure of the liquid helium
storage.
Therefore, low energy $p$'s were out of the BESS
acceptance region in this beam experiment due to the opposite
deflection. 

\section{Analysis}
\label{sec:anal}

\subsection{Beam identification}

To determine $N_{\rm inc}$ the beam line detectors must unambiguously
identify the incident particle and precisely determine incident
energy, position, and angle.
The incident kinetic energy, $E_{\rm TOI}$, 
was derived from $\beta_{\rm T4-T3}$, 
where $\beta$, the velocity of incident particle, was obtained from a
pair of timing measurements.
Note that (i) less accurate energy resolution was obtained from the
deflection measurement due to the multiple scattering and limited
track 
length, and (ii) it is not appropriate to derive the incident energy
from the $\beta_{\rm T4-T1}$, etc. due to the energy losses in the
beam line in spite of the much better $1/\beta$ resolution.
The accuracy of energy determination was 1~\% in the very low energy
region around 0.2~GeV.
Around 1~GeV, energy determination is obtained with 4~\% accuracy due
to the constant $1/\beta_{\rm T4-T3}$ resolution.
Systematic error of the absolute energy is estimated to be $\pm$
1~\% due to calibration of the time of flight measurements and energy
losses in the beam line.
The beam trajectory obtained by DC1 and DC2 is extrapolated to the
BESS detector taking into account the fringing field of the
solenoid. 
The accuracy of the incident position and angle around 1~GeV
were 2.5 mm and 2 mrad (rms deviation), respectively. 

\begin{figure}[hbt]
  \begin{center}
    \includegraphics[width=13.5cm]{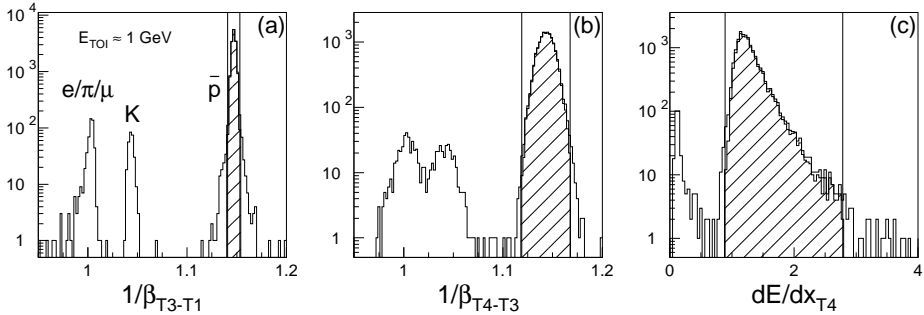}
  \end{center}
  \caption  {$1/\beta$ and d$E$/d$x$ cut by trigger counters for
    $\bar{p}$'s of $E_{\rm TOI} \sim$ 1.0~GeV. $1/\beta_{\rm
    T3-T1}$, $1/\beta_{\rm T4-T3}$, and d$E$/d$x_{\rm T4}$ are shown
    in (a), (b), and (c), respectively. The hatched histogram
    indicates events survived after applying the $\bar{p}$ beam
    selection.} 
  \label{fig:id}
\end{figure}

To identify $\bar{p}$'s and $p$'s, 
we required that the energy losses in T1 -- T4 and the velocities
measured by various combination of trigger counters are consistent
with $p$'s.
Figure~\ref{fig:id} shows examples of these cuts for
$\bar{p}$'s of $E_{\rm TOI} \sim$ 1~GeV.
Since $1/\beta$ distribution shows a clear separation between
$\bar{p}$, kaon, and pion/muon/electron particles which allows
incident beam particles to be unambiguously identified.   
To ensure that incident particles arrive at the top of the BESS
instrument, 
we further required (i) at least one hit in
the upper TOF hodoscope and (ii) beam trajectory agreement with the
upper TOF hit position in both $r$-$\phi$ and $y$-$z$ planes, 
where the hit point in TOF is obtained from the segmentation of
hodoscope in $\phi$ direction and timing difference of PMT at the both
end in $z$ direction.
These two requirements guaranteed that the incident particles pass
through T4 and AC without large angle scattering or interaction,
resulting in the precise determination of $N_{\rm inc}$.

In order to compare the beam data and BESS MC results, a MC data
set was generated as follows. 
To estimate interaction and
energy losses in T4 and AC, which were located just upstream of BESS,
they were described in the BESS MC.
Input kinematics of beam particles was obtained from beam data
event by event.
This allows comparisons between beam data and MC results under the
same conditions. 
Data sets are referred to as \beam\  and \mc\   samples, respectively.

\subsection{Detection efficiency}
\label{sec:eff}
In order to select non-interacting $\bar{p}$'s, we applied the same
cuts as those used in the standard $\bar{p}$ analysis for the flight
data~\cite{kn:pbar98}:
\begin{enumerate}
\item select events with a single downward-traveling particle fully
contained in the fiducial region of the tracking detectors; 
\item require only one (one or two) hit in the upper (lower) TOF
hodoscopes \cite{kn:tofl};
\item require the TOF hit position consistent with the extrapolation
of the trajectory determined by JET and IDCs (not by beam line
trackers); 
\item require that d$E$/d$x$ measurements of upper and lower TOF are
consistent with those of $p$'s;
\item require a quality track in terms of fitting chi-squares, number
of hits used in fitting, and etc.;
\item require that the mass derived from momentum and velocity
measurements is consistent with $p$ mass.
\end{enumerate}
Cuts (1)--(4) reject most of the events with interactions. 
Distributions of the cut parameters are provided in
Ref.~\cite{kn:d-asa}.
Cut (5) assures the quality of momentum and velocity measurements. 
Cut (6) is to identify $\bar{p}$'s. 
Selection criteria for $p$'s are identical to the above except for
the sign of momentum. 
The efficiencies of cuts (1)--(4) are different for $p$'s and
$\bar{p}$'s because of the different interaction cross sections.
However, the efficiencies of cuts (5) and (6) are the same for $p$'s
and $\bar{p}$'s due to the symmetrical configuration of the detector.
Therefore, we express the detection efficiency ($\varepsilon$) as the
product of the non-interacting efficiency ($\varepsilon_{\rm
non-int}$) associated to the cuts (1)--(4) and the quality and
particle-identification efficiency ($\varepsilon_{\rm Q-ID}$)
associated to cuts (5) and (6):
\[ \varepsilon = \varepsilon_{\rm non-int} \cdot \varepsilon_{\rm
Q-ID}.\] 
Since $\varepsilon_{\rm non-int}$ relies on the \mc\  calculation and
$\varepsilon_{\rm Q-ID}$ can be estimated by using the $p$ sample in
the flight data, we should verify the former efficiency by the beam
experiment.

\subsection{Beam-related corrections and systematic errors}
\label{sec:err}
The following beam related corrections are applied to
$\varepsilon_{\rm non-int}$ and the systematic errors are estimated.
\begin{enumerate}
\item Beam dump effect:
Annihilation of $\bar{p}$'s at the beam dump located 3~m downstream of 
the BESS detector produces secondary particles, some of which generate
delayed hits in the TOF counters. 
After eliminating the beam dump effect by removing the delayed hits
from the hits in the upper and lower TOF hodoscopes, the remaining
correction was negligible. 
The systematic uncertainty was estimated to be 0.002.

\item Accidental tracks:
Multi-track events without a vertex in the detector were considered to
be a beam-related accidental track and the track is removed from
the event. 
However, an accidental track would be mis-identified as an interaction
if the accidental particle is close to the incident $\bar{p}$. 
The remaining correction for mis-identification was
0.001 -- 0.005 and the systematic uncertainty of the 
$\varepsilon_{\rm non-int}$ was estimated to be 0.001.
The correction and its error for $p$'s were negligible because
of the smaller incident beam intensity.

\item Interaction at upper TOF:
The use of upper TOF information to identify the non-interacting beam
entering the BESS detector rejects a small fraction of events with
interaction inside BESS, which underestimates $N_{\rm inc}$ and thus 
overestimates $\varepsilon_{\rm non-int}$. 
The correction for this effect was estimated by
varying the cut parameters of the selections and was estimated to be
$-$0.01 -- $-$0.005. 
The rms deviation of the difference in $\varepsilon_{\rm non-int}$
between \beam\  and \mc\  results was taken as the systematic
uncertainty, which was evaluated to be 0.005.

\item Beam energy:
Since the systematic uncertainty of the beam energy is $\pm$1~\%, the
corresponding uncertainty in $\varepsilon_{\rm non-int}$ is negligible for 
energies above 0.16~GeV.
However, it is difficult to reliably estimate the corresponding
systematic errors in $\varepsilon_{\rm non-int}$ derived from \beam\
below 0.16~GeV, where $\varepsilon_{\rm non-int}$ changes rapidly with
energy.
\end{enumerate}

In total, beam related systematic uncertainty was estimated to be less
than 0.01 for energies up to 1~GeV except for very low energy region
where the efficiency rapidly changes with energy.

\section{Results}
\label{sec:rslt}

\begin{figure}[hbt]
  \begin{center}
    \includegraphics[width=13.5cm]{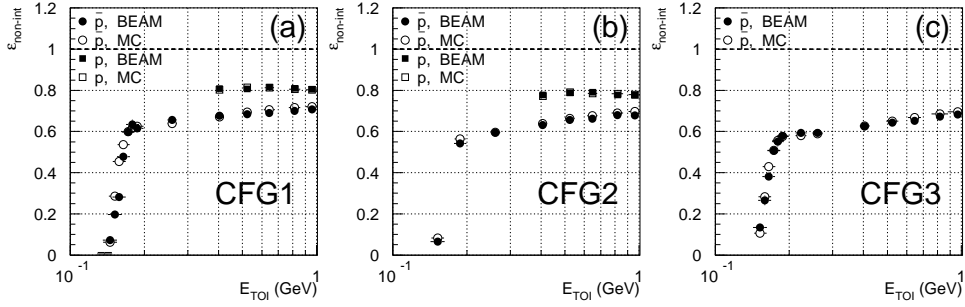}
  \end{center}
  \caption  {Direct measurement of $\varepsilon_{\rm non-int}$ for
  $\bar{p}$'s and $p$'s in (a) \cfg{1}, (b) \cfg{2}, and (c)
  \cfg{3}.
  The error bars include both statistical and systematic
  uncertainties, while they are too small to see in this figure 
  (see Fig.~\ref{fig:effcmp}, instead).
   }
\label{fig:effdir}
\end{figure}

Figure~\ref{fig:effdir} shows $\varepsilon_{\rm non-int}$ for
$\bar{p}$'s and $p$'s for each configuration derived from
\beam\  and \mc\  samples. 
Reflecting the energy dependence of cross sections, $\varepsilon_{\rm
non-int}$ for $\bar{p}$'s gradually decreases with decreasing energy,
while $\varepsilon_{\rm non-int}$ for $p$'s shows little dependence on 
energy. 
Among \beam\ data, the differences between \cfg{1--3} in
$\varepsilon_{\rm non-int}$ are 7, 5, and 3~\% at
energies of 0.2, 0.4, and 1~GeV, respectively. 
Although large discrepancies exist below 0.16~GeV
where $\varepsilon_{\rm non-int}$ rapidly drops due to stopping of
incident $\bar{p}$'s in the instrument,
they can be explained by the systematic uncertainty of 1~\% in the
beam energy determination. 

\begin{figure}[hbt]
  \begin{center}
    \includegraphics[width=7.0cm]{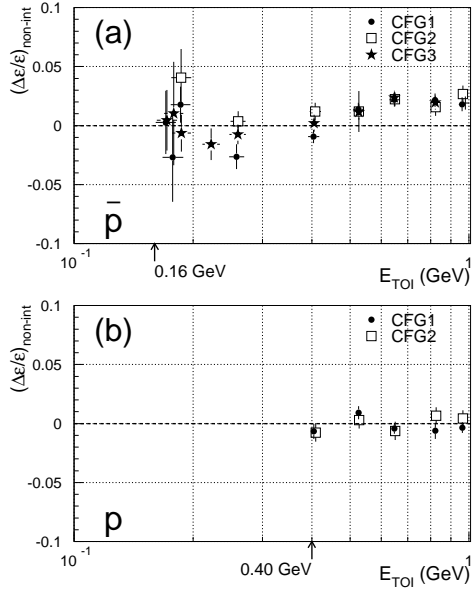}
  \end{center}
  \caption {Relative difference in $\varepsilon_{\rm non-int}$ of \mc\
  from \beam\  for (a) $\bar{p}$'s and (b) $p$'s.}
  \label{fig:effcmp}
\end{figure}

The relative differences in $\varepsilon_{\rm non-int}$ between \beam\
and \mc\  samples ($(\Delta \varepsilon / \varepsilon)_{\rm non-int} = 
(\varepsilon_{\rm MC} -\varepsilon_{\rm beam}) / \varepsilon_{\rm
beam}$) are shown in Fig.~\ref{fig:effcmp} from 0.16 to 1.0~GeV for  
$\bar{p}$'s and 0.4 to 1.0~GeV for $p$'s, where beam related
systematic errors were kept low and well estimated.
As shown, $(\Delta \varepsilon / \varepsilon)_{\rm non-int}$
was kept within $\pm$5~\% for $\bar{p}$'s, 
demonstrating good agreement between \beam\  and \mc\  samples. 
Above 0.5~GeV small but significant discrepancies (2~\%) in 
$(\Delta \varepsilon / \varepsilon)_{\rm non-int}$ are present,
while statistical accuracy of \beam\  data limited the discussion 
in the low energy region below 0.3~GeV.
For $p$'s, $(\Delta \varepsilon / \varepsilon)_{\rm non-int}$
was kept within $\pm$2~\%.
Moreover, good reproducibility of $(\Delta \varepsilon /
\varepsilon)_{\rm non-int}$ are obtained between different 
configurations, {\it i.e.}, 
the maximum deviation of $\Delta \varepsilon_{\rm non-int}$ between
all configurations are close to the statistical accuracy of each data
point for all energies (mostly $\sim$ 0.005). 
Note that \cfg{1--3} were selected to represent the typical incidence of
cosmic-ray particles, and the properties of non-interacting particles
show very good agreement between \beam\  and \mc\  samples. 

As a result, 
these results clearly demonstrates that above 0.16~GeV (for $p$'s
above 0.4~GeV) systematic error in $\varepsilon_{\rm non-int}$ is
determined to be within $\pm$5~\% ($\pm$2~\%).
Since systematic error in $\varepsilon_{\rm Q-ID}$ was well estimated
to be within $\pm$1~\% using unbiased $p$ sample in flight data, 
systematic error in $\varepsilon_{\rm non-int}$ can be considered as
that in detection efficiency. 

\section{Discussion}
\label{sec:dsc}
In a detailed study of the interaction processes in the BESS
MC~\cite{kn:d-asa}, we found no significant difference in
$\varepsilon_{\rm non-int}$ for different hadronic packages, {\sc
gheisha} and {\sc fluka}~\cite{kn:fluka} (implemented in {\sc
geant}{\footnotesize 3}, denoted as {\sc gfluka}) 
if we assume the same total inelastic cross section.
This means that the multiplicity and angular distribution of secondary
particles are less important as far as $\varepsilon_{\rm non-int}$ is
concerned. 
Any interactions in the BESS detector are well discriminated by a
combination of various and independent measurements. 
The only factors relevant to the efficiency are essentially the total
inelastic cross section and the material amount. 
We studied d$E$/d$x$ of stopping cosmic rays using the flight data
and compared it with the BESS MC simulation.
As a result we confirmed that the amount of material was correctly
implemented in the BESS MC with an accuracy of $\pm$0.5~g/cm$^2$. 

Since $\sim$1/3 of the incident particles interact in the instrument,
the $\pm$5~\% agreement between the measured and calculated
efficiencies means that the previous measurements of $\bar{p}$-nuclei
inelastic cross sections were verified at least within an accuracy of
about $\pm$15~\%. 

\begin{figure}[hbt]
  \begin{center}
    \includegraphics[width=8.0cm]{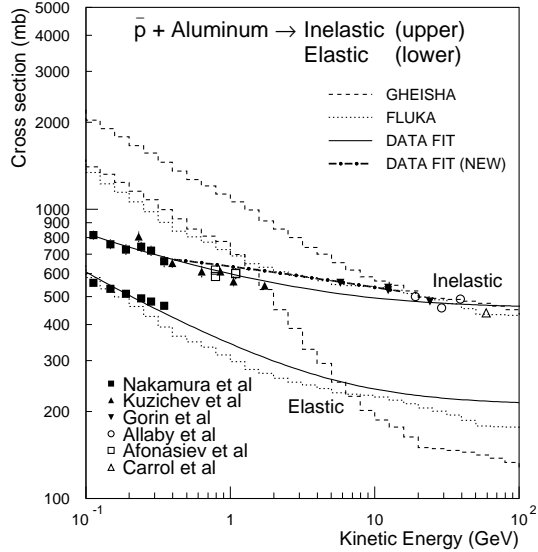}
  \end{center}
  \caption {Inelastic and elastic cross sections for $\bar{p}$ to
  aluminum. Histograms are from the {\sc gheisha} and {\sc fluka} codes
  and data points~\cite{kn:pbcprl} from Nakamura \etal\  (filled
    squares); Kuzichev \etal\  
  (filled triangles); Gorin \etal\  (inverted triangles); Allaby \etal\
  (open circles); Afonas'ev \etal\  (open squares); and Carroll \etal\
  (open triangles). The solid curves are fit with these data points and
  used in the BESS MC. The dash-dotted curve is a new fit that better
  reproduces the results of the beam experiment.}   
  \label{fig:cs}
\end{figure}

Figure~\ref{fig:cs} shows the inelastic and elastic cross sections for
$\bar{p}$ to aluminum used in the BESS MC ({\sc gheisha(bess)}, solid
curve), as well as those of the original {\sc gheisha} code, {\sc
fluka} code, and experimental data~\cite{kn:pbcprl}.
The data of Kuzichev \etal\  only represent annihilation cross
section. 
Assuming that the non-annihilation inelastic cross section is about
10~\% of the total cross section as assumed in the {\sc gheisha} code,
we modified the data of Kuzichev \etal\  and refit all the data below
15~GeV, indicated by the dash-dotted curve in Fig.~\ref{fig:cs}
({\sc gheisha(new)}). 
With this new cross section, $\varepsilon_{\rm non-int}$ was
reproduced with better accuracy up to 1~GeV.
As a result, the systematic error in detection efficiency can be
further reduced to $\pm$2~\% from 0.3 to 1~GeV for $\bar{p}$'s.

Since we do not have the measured data of $\varepsilon_{\rm non-int}$ 
above 1~GeV, the systematic error in this energy region was estimated
from the relative difference in $\varepsilon_{\rm non-int}$ among
several hadronic packages ({\sc gheisha(new)}, {\sc gheisha(bess)},
{\sc gfluka}).
The systematic error in detection efficiency above 1~GeV was estimated
to be $\pm$5~\%. 

\section{Conclusion}
\label{sec:conc}
Through an accelerator beam experiment, measurements of the 
detection efficiency for $\bar{p}$'s and $p$'s in BESS were performed
below 1~GeV.
The measurements and following studies indicated that the detection
efficiency was mainly determined from material amount in the
instrument and cross sections with nuclei, essentially regardless of
the simulation details such as multiplicity and angular distribution
of secondaries. 
As a result, we confirmed that the relative systematic error in
detection efficiency derived from the simulation was kept below 5\%
and 2\% for $\bar{p}$'s and $p$'s, respectively.
It can be applied to systematic errors in cosmic-ray
$\bar{p}$ measurements \cite{kn:pbar95,kn:pbar97,kn:pbar98,kn:pbar00},
as well as future BESS experiments including high statistics
long-duration flights\cite{kn:polar} providing the instrumental
features of BESS are maintained.  
Moreover, we found the total advantage of cylindrical configuration
with superconducting solenoid by eliminating possible disadvantage in
determining detection efficiencies accompanied with 30--40~\%
$\bar{p}$ interaction losses in the instrument.
By increasing the reliability of the cosmic-ray $\bar{p}$ spectrum,
these results will enable us to carry out the most sensitive-ever
investigation on the origin of cosmic-ray $\bar{p}$'s.

\begin{ack}
We appreciate support during accelerator beam experiments by
Mr. M. Taino, Dr. M. Ieiri (KEK), Dr. A. Ichikawa (Kyoto Univ.),
Mr. H. Takahashi (Kyoto Univ.), and other members of E373 group.
We also thank Mr. S. Tsuno (Tsukuba Univ.) and Dr. H. Iwasaki (KEK)
for allowing use of their drift chambers (DC1, DC2).
This work was supported by a Grant-in-Aid for Scientific Research from
the Japanese Ministry of Education, Science and Culture.
Analysis was performed using the computing facilities at ICEPP, the
University of Tokyo. 
Some of the authors were supported by Japan Society for the Promotion
of Science.
\end{ack}


\begin{thebibliography}{99}
\bibitem{p-nuclei} Note that $p$-nuclei cross sections below 1~GeV
    in the BESS MC are also modified from original {\sc gheisha} code,
    based on the LaRC model calculation~\cite{kn:larc}.
\bibitem{kn:tofl} The different selection for lower TOF hodoscope is
    applied to save events with delta ray.
\end{thebibliography}
\end{document}